\documentclass[conference]{IEEEtran}
\IEEEoverridecommandlockouts

\usepackage{cite}
\usepackage{amsmath,amssymb,amsfonts}
\usepackage{algpseudocode} 
\usepackage{algorithm}     
\usepackage{graphicx}
\usepackage{textcomp}
\usepackage{color,soul}
\usepackage{lipsum} 
\usepackage{xurl}

\usepackage{booktabs} 
\usepackage{array} 
\usepackage{makecell} 
\usepackage{url}
\usepackage{booktabs}
\usepackage{amssymb}
\usepackage{tabularx}
\usepackage{optidef}
\usepackage{graphicx}
\usepackage{wrapfig}
\usepackage{afterpage}
\usepackage{float}
\usepackage{multirow}
\usepackage{array}
\usepackage{hhline}
\usepackage{arydshln}
\usepackage{tabularx}
\usepackage{makecell}
\usepackage{subcaption}
\usepackage{booktabs}
\usepackage{indentfirst}
\usepackage{adjustbox}
\usepackage{amsmath}
\usepackage{caption}
\usepackage{soul}
\usepackage{soulpos}

\DeclareMathOperator*{\argmin}{arg\,min}

\usepackage{bm} 
\usepackage{mathptmx} 

\usepackage{ifthen} 
\usepackage{xcolor} 
\usepackage{ulem}   

\newcommand{\showchanges}{0}

\newcommand{\added}[1]{\ifthenelse{\equal{\showchanges}{1}}{\textcolor{blue}{#1}}{#1}}
\newcommand{\deleted}[1]{\ifthenelse{\equal{\showchanges}{1}}{\textcolor{red}{\sout{#1}}}{}}

\usepackage{bm}
\begin{document}

\title{Adversarial Robustness of Traffic Classification under Resource Constraints: Input Structure Matters%
\thanks{This work was partially supported by project SERICS (PE00000014) under the MUR National Recovery and Resilience Plan funded by the European Union - NextGenerationEU.}
}
\author{
    \IEEEauthorblockN{Adel Chehade, Edoardo Ragusa, Paolo Gastaldo, and Rodolfo Zunino}
    \IEEEauthorblockA{
        DITEN, University of Genoa, Italy \\
        Email: adel.chehade@edu.unige.it, \{edoardo.ragusa, paolo.gastaldo, rodolfo.zunino\}@unige.it
    }
}

\maketitle


\begin{abstract}
Traffic classification (TC) plays a critical role in cybersecurity, particularly in IoT and embedded contexts, where inspection must often occur locally under tight hardware constraints. We use hardware-aware neural architecture search (HW-NAS) to derive lightweight TC models that are accurate, efficient, and deployable on edge platforms. Two input formats are considered: a flattened byte sequence and a 2D packet-wise time series; we examine how input structure affects adversarial vulnerability when using resource-constrained models. Robustness is assessed against white-box attacks, specifically Fast Gradient Sign Method (FGSM) and Projected Gradient Descent (PGD). On USTC-TFC2016, both HW-NAS models achieve over 99\% clean-data accuracy while remaining within 65k parameters and 2M FLOPs. Yet under perturbations of strength 0.1, their robustness diverges: the flat model retains over 85\% accuracy, while the time-series variant drops below 35\%. Adversarial fine-tuning delivers robust gains, with flat-input accuracy exceeding 96\% and the time-series variant recovering over 60 percentage points in robustness, all without compromising efficiency. The results underscore how input structure influences adversarial vulnerability, and show that even compact, resource-efficient models can attain strong robustness, supporting their practical deployment in secure edge-based TC.
\end{abstract}

\begin{IEEEkeywords}
Traffic classification, efficient deep models, adversarial robustness, neural architecture search, edge intelligence
\end{IEEEkeywords}

\section{Introduction}

Traffic classification (TC) is a foundational task in network monitoring, essential for malware detection, policy enforcement, and intrusion prevention~\cite{wang2017malware, zhu2023cmtsnn}. In modern settings, traffic is often encrypted or obfuscated, limiting the effectiveness of traditional approaches that rely on protocol inspection or hand-crafted features \cite{survey_DONG2024128444}. Deep neural networks (DNNs) have shown strong performance on raw traffic data, using byte sequences directly extracted from network packets~\cite{wang2017end, wang2017malware, nas_zhang2023automatic}; this approach bypasses costly feature engineering and adapts well to evolving traffic patterns. 
This work examines how compact models operating on raw byte-level flow inputs behave under adversarial perturbations, and how their robustness is influenced by input structure; it also investigates whether such models, targeted for deployment on sensing nodes and edge gateways, can be made robust despite their limited capacity.


While effective, most DNN-based classifiers are still too resource-intensive for practical use on embedded platforms \cite{our_chehade2025iscc}. Their high memory requirements, inference latency, and computational load make them difficult to run on low-power devices. Yet, performing classification locally is often preferred over cloud offloading, as it reduces end-to-end latency, enhances privacy, and avoids reliance on network connectivity~\cite{ragusa2024combining, our_chehade2025iscc}. Supporting this shift toward edge intelligence requires models optimized not only for accuracy but also for hardware efficiency. 
To this end, we adopt a hardware-aware neural architecture search (HW-NAS), which is becoming a key technology for building efficient models tailored to constrained platforms; it automatically explores candidate architectures and selects those that meet strict limits on parameters, computation, and memory \cite{our_chehade2025tnsm}.

Meanwhile, DNNs are known to be vulnerable to adversarial examples, which are small, structured perturbations crafted to cause misclassification \cite{roshan2024untargeted, alhussien2024constraining}. This is especially problematic in TC, where models operate in security-sensitive contexts \cite{zhang2024adversarial}. Attackers with access to model parameters and gradients (white-box) can craft targeted perturbations; black-box adversaries may rely on queries or surrogate models. Unlike image data, such perturbations in network traffic are not visually interpretable; instead, they subtly disrupt byte-level features the model relies on for classification. Yet, adversarial robustness in TC under hardware constraints remains underexplored, and this study provides a focused analysis in that direction.

We investigate the effect of input representation on adversarial robustness, and whether architectures optimized for efficiency can still provide resilience without sacrificing deployability. Our evaluation focuses on adversarial attacks constrained in the $\ell_\infty$ norm, which limit the maximum allowable change to any input feature. Specifically, we consider the Fast Gradient Sign Method (FGSM)~\cite{fgsm_goodfellow2014explaining} and Projected Gradient Descent (PGD)~\cite{pgd_madry2017towards}; we perform a post-hoc robustness evaluation of HW-NAS-optimized models, followed by lightweight adversarial fine-tuning to assess potential improvements.

\noindent Our main contributions are as follows:
\begin{itemize}
    
    \item We test the adversarial robustness of compact DNNs derived via HW-NAS, proving that, even under strict efficiency constraints, robustness can still be induced through fine-tuning.


    \item We study the effect of input structure on adversarial vulnerability by comparing two widely used traffic representations: flattened byte sequences and 2D packet-wise time series. The results demonstrate that the choice of input structure can significantly affect model robustness.

    \item On USTC-TFC2016~\cite{wang2017malware}, both input formats yield high accuracy on clean data, with 99.60\% for flat and 99.18\% for time-series input. While performance is comparable to the state of the art, our models achieve this with much lower complexity, using less than 65k parameters, 2M FLOPs, and 5k maximum intermediate tensor size.

   \item We assess the robustness of both models under white-box adversarial attacks using FGSM and PGD across a range of perturbation strengths. Without any defense, the flat-input model retains up to 86.49\% (FGSM) and 74.78\% (PGD), while the time-series model 
   exhibits far lower robustness, dropping to 32.23\% and 23.01\%, respectively.

    \item We apply batch adversarial training with FGSM. After adaptation, the flat-input model exceeds 96\% accuracy under FGSM and 93\% under PGD. The time-series model, while still more sensitive, improves markedly, reaching over 88\% and 84\%. On clean data, both models retain high accuracy at 98.98\% and 98.71\%, respectively.

\end{itemize}

\section{Related Work}
\subsection{Traffic Classification and Input Representations}

Early TC relied on port-based heuristics and payload inspection. Port numbers became unreliable due to tunneling, encryption, and dynamic allocation~\cite{ports_niu2019heuristic,survey_DONG2024128444}. Deep packet inspection (DPI), while effective on plaintext, raises privacy concerns and performs poorly on encrypted traffic~\cite{signature_dpi1_wang2020automatic,dpi_hongke2022dpi}.

Modern approaches adopt learning-based models at multiple granularities, including packet (individual transmission units), flow (unidirectional sequences), and session (bidirectional flow pairs)~\cite{our_chehade2025iscc}. Classical ML pipelines often use statistical features extracted from flows~\cite{survey_DONG2024128444}, but their reliance on handcrafted inputs limits adaptability.

Deep learning (DL) models have increasingly been applied to raw byte sequences from network traffic. A common strategy flattens each session or packet into a fixed-length vector, typically processed by 1D-CNNs~\cite{wang2017end,wang2017malware,our_chehade2025tnsm}. This view captures local byte-level patterns but ignores packet structure; it was initially used for encrypted traffic and malware classification~\cite{wang2017end,wang2017malware}, and later extended to more complex designs. For example, L2-BiTCN-CNN~\cite{li2024l2} incorporates hierarchical temporal features and 2D convolutions. Flow-level methods include cascaded dense networks~\cite{xu2024cascaded} or ConvNet-based confidence estimation~\cite{li2025trustworthy}. Packet-level architectures range from CNN–RNN hybrids using raw bytes~\cite{ieeeaccess_wang2020deep} to deeper 2D-CNN models designed for structured packet matrices~\cite{yu2025model}.

Alternatively, some studies adopt a time-series perspective, segmenting flows into fixed numbers of packets and representing each packet as a byte vector (e.g., 10×1000 bytes)~\cite{yao2019identification,zhu2023cmtsnn,yao2019capsule}. This format preserves spatial and temporal order, allowing sequential models such as CNN–RNN hybrids or capsule-based architectures to learn cross-packet dependencies. At the same time, these inputs increase model complexity and resource usage during training and inference.

Most existing studies emphasize accuracy; few consider how input representation influences adversarial robustness in deep traffic classifiers, which is the core aspect of this work.

\subsection{Neural Architecture Search under Hardware Constraints}

Neural Architecture Search (NAS) has been widely used to automate the design of neural networks in domains like vision and language~\cite{1000papersNAS,chitty2023neural_ieee-access}, but its use in TC remains limited. NAS has also been applied to malware detection~\cite{nas_zhang2023automatic}, where it yields strong results under accuracy-driven objectives. These results highlight the flexibility of NAS, but most approaches focus solely on accuracy and overlook deployment feasibility.

HW-NAS extends this by optimizing architectures under resource limits such as parameter count, FLOPs, and memory usage~\cite{ragusa2024compression,ragusa2024combining}. Prior studies have demonstrated its effectiveness for session-level classification~\cite{our_chehade2024tiny,our_chehade2025tnsm}, including deployment on microcontrollers~\cite{our_chehade2025iscc}. Existing HW-NAS models are typically evaluated only on clean data, raising questions about their robustness and how fine-tuning might improve their resilience to adversarial perturbations.

\subsection{Adversarial Robustness in Traffic Classification}

DL models  have shown strong performance in TC but remain vulnerable to structured input perturbations. Early work evaluated gradient-based evasion attacks such as FGSM and PGD (both under $\ell_\infty$ constraints), DeepFool (minimizing $\ell_2$ distortion), and the Jacobian-based Saliency Map Attack (JSMA), which alters a few input features based on saliency~\cite{maarouf2021evaluating,roshan2024untargeted}. Some works also examined robustness improvements via adversarial training, Gaussian noise augmentation, or confidence-based filtering~\cite{zhang2024adversarial,roshan2024untargeted}.

Other studies examined structured features derived from flow metadata or protocol fields, where attacks must preserve domain semantics (e.g., timing, ports, flow statistics), making robustness highly sensitive to the input format and model design~\cite{alhussien2024constraining}. In contrast, evaluations on raw byte inputs across packets, flows, and time-series formats show that robustness varies significantly with traffic encoding~\cite{sadeghzadeh2021adversarial}.

Yet, most prior work assumes large, server-grade models. The behavior of compact, resource-efficient classifiers under adversarial conditions remains underexplored. We address this gap by evaluating HW-NAS–optimized models against $\ell_\infty$-bounded white-box attacks that have been shown to challenge DL models in TC~\cite{roshan2024untargeted}. We also apply lightweight adversarial fine-tuning as a defense within deployment constraints.

\section{Methodology}

We aim to make lightweight models robust against adversarial attacks while remaining suitable for deployment. To this end, we identify three key design aspects that influence robustness under strict efficiency constraints: (i) input representation, (ii) architecture design and search under hardware limits, and (iii) post-hoc adversarial analysis and adaptation. The following sections detail each of these points.
\subsection{Input Representation}

We operate at the flow level, where a flow is a unidirectional packet sequence defined by a 5-tuple: source IP, destination IP, source port, destination port, and protocol. Packets are treated as raw byte streams representing transmitted network data.

Each flow is encoded using one of two common strategies. The first treats the entire flow as a flattened byte sequence, normalized to a fixed length via truncation or padding, as adopted in prior works \cite{wang2017end, wang2017malware}. The second preserves packet structure by encoding each flow as a 2D time series matrix with shape $N \times M$, where $N$ is the number of packets and $M$ the bytes per packet. This view captures spatial-temporal structure more explicitly, as explored in \cite{yao2019identification, zhu2023cmtsnn, yao2019capsule}; 
this distinction can shape how perturbations propagate across local or temporal patterns in each format.


\subsection{Designing Compact Models under Hardware Constraints}

\subsubsection{Problem Formulation}

We formalize architecture search as a constrained optimization problem that balances predictive accuracy with resource usage. Let \( \mathcal{A} \) denote the set of candidate architectures, and \( a \in \mathcal{A} \) a specific architecture. We adopt a HW-NAS framework, where the goal is to approximate a solution \( a^* \in \mathcal{A} \) that maximizes validation accuracy while satisfying hardware constraints. The problem is defined as:

\begin{maxi}|s|
    {a \in \mathcal{A}}{\text{Accuracy}_{\text{val}}(w^*(a), a)}
    {}{}
    \addConstraint{w^*(a) = \argmin_w \mathcal{L}_{\text{train}}(w, a)}
    \addConstraint{|P(a)| < F_{\text{Th}}}     
    \addConstraint{|T(a)| < R_{\text{Th}}}     
    \addConstraint{\text{FLOPs}(a) < \text{FLOPs}_{\text{Th}}}
    \label{eq:hw_nas_opt}
\end{maxi}

Here, \( w^*(a) \) denotes the optimal weights for architecture \( a \), minimizing training loss. The constraints reflect hardware limits: \( |P(a)| \) (bounded by \( F_{\text{Th}} \)) controls the number of trainable parameters (flash), \( |T(a)| \) (bounded by \( R_{\text{Th}} \)) limits the maximum intermediate tensor size (RAM), and FLOPs (bounded by \( \text{FLOPs}_{\text{Th}} \)) restrict the per-inference computational cost. These thresholds are chosen to support deployment on highly constrained platforms, including microcontrollers, consistent with embedded NAS studies~\cite{ragusa2024combining, ragusa2024compression,our_chehade2025iscc}.

\subsubsection{Search Space and Architecture Design}

Motivated by prior findings that 1D CNNs offer strong accuracy for TC tasks~\cite{wang2017end, wang2017malware, our_chehade2025tnsm} and are efficient under hardware limits~\cite{our_chehade2024tiny, our_chehade2025iscc}, we construct a search space composed of modular 1D convolutional blocks. CNNs are well-suited to both input formats considered in this work, though in different ways: they extract local patterns from flattened byte sequences and capture temporal structure in the time-series matrices, without the complexity of recurrent or transformer-based models.

Each block includes a Conv1D layer with tunable filter count, kernel size, stride, and padding, followed by batch normalization and ReLU activation. Pooling is applied using either max or average operations, and dropout is included for regularization. A global average pooling (GAP) layer comes before the final dense classifier. The number of blocks is variable, and all hyperparameters are selected dynamically during the search to fit task needs and hardware limits.

\subsubsection{Evolutionary NAS Algorithm}

We adopt an evolutionary search strategy, which has proven effective for hardware-aware architecture optimization~\cite{ragusa2024compression, our_chehade2025iscc}. The algorithm begins with an initial parent architecture $a_0 \in \mathcal{A}$, and proceeds for $N_g$ generations. At each generation $g$, a set of $N_c$ child architectures is generated by applying random mutations to the current parent $a_p$.

The mutation operator $R_m(a_p)$ either modifies existing blocks (e.g., adjusting filter count or kernel size), adds new blocks with randomly sampled hyperparameters, or removes blocks entirely.
Each mutated child $a_c$ is evaluated for feasibility: architectures that exceed predefined  hardware thresholds are discarded without training. Mutation continues until the required number of valid children is reached. Feasible candidates are then trained on the training set $\mathcal{X_T}$ and evaluated on a validation set $\mathcal{X_V}$. Among all valid candidates in the current generation, the one achieving the highest validation accuracy becomes the new parent for the next generation.

This process iteratively refines the architecture, promoting gradual improvements while ensuring hardware compliance. At the end of the search, the final output $a^*$ is chosen as the hardware-feasible architecture that achieved the highest validation accuracy across all generations.


\subsection{Post-Hoc Robustness Analysis}

\subsubsection{White-Box Adversarial Evaluation}

We assess the robustness of the final HW-NAS architectures through post-hoc adversarial analysis on the two input formats used in this study: the flattened 1D byte sequence and the 2D time-series matrix. This evaluation is conducted under a white-box setting, where the attacker has full access to model parameters and gradients; 
the goal is to examine how input format affects vulnerability under similar hardware budgets.

We focus on adversarial attacks constrained in the $\ell_\infty$ norm, namely FGSM and PGD, which are widely used in robustness evaluation. These attacks limit the maximum change allowed per input feature, making them well-suited for evaluating robustness to small, bounded perturbations typical in adversarial settings. More complex methods such as C\&W and DeepFool, which rely on different norm constraints and incur higher computational cost, are excluded from this study.

The analysis begins with FGSM, a single-step method that perturbs inputs in the direction of the loss gradient:

\begin{equation}
    x^{\text{adv}} = x + \epsilon \cdot \text{sign} \left( \nabla_x \mathcal{L}(f(x), y) \right)
\end{equation}

Here, $f(x)$ denotes the model output, $\mathcal{L}$ is the loss function (e.g., cross-entropy), $y$ is the true label, and $\epsilon > 0$ is the perturbation budget. The gradient $\nabla_x \mathcal{L}(f(x), y)$ points toward the direction that most increases the loss, producing minimal but effective perturbations bounded by the $\ell_\infty$ norm.

To assess robustness under stronger perturbations, we apply PGD, a multistep variant of FGSM that iteratively updates the input via gradient ascent on the loss, followed by projection onto the $\ell_\infty$ ball of radius $\epsilon$ centered at the original input $x$. Formally, the update rule is defined as:

\begin{equation}
    x_0^{\text{adv}} = x, \quad
    x_{t+1}^{\text{adv}} = \Pi_\epsilon^x \left( x_t^{\text{adv}} + \alpha \cdot \text{sign} \left( \nabla_{x_t^{\text{adv}}} \mathcal{L}(f(x_t^{\text{adv}}), y) \right) \right)
\end{equation}

where $\alpha$ is the step size, $t$ is the iteration index, $\mathcal{L}(f(x), y)$ denotes the loss function between the model prediction $f(x)$ and the true label $y$, and $\Pi_\epsilon^x(\cdot)$ denotes projection onto the $\ell_\infty$ ball of radius $\epsilon$ centered at $x$.

\subsubsection{Fine-Tuning as Defensive Adaptation}
Adversarial robustness is targeted via fine-tuning using FGSM in a batch-level training regime. Due to its efficiency and low overhead, FGSM is well suited for on-the-fly adversarial augmentation during training. In each mini-batch, a subset of training inputs is perturbed via FGSM, and the model is optimized jointly on both clean and adversarial examples. The training loss is:

\begin{equation}
    \min_{h} \sum_{i \in \mathcal{C}} \left[ \mathcal{L}(f(x_i; h), y_i) + \mathcal{L}(f(x_i^{\text{adv}}; h), y_i) \right]
\end{equation}

Here, $x_i$ and $x_i^{\text{adv}}$ are clean and FGSM-perturbed inputs, $y_i$ is the corresponding label, $\mathcal{C}$ is their shared index set, and $h$ denotes the model parameters. After fine-tuning, models are re-evaluated under FGSM and PGD to quantify robustness gains across input formats and perturbation strengths.

\section{Experimental Setup}
\subsection{Dataset and Preprocessing} \label{sec:dataset}

We conduct experiments on the USTC-TFC2016 dataset~\cite{wang2017malware}, which contains labeled PCAP traffic from both benign applications and malware families.
The dataset contains 734{,}168 flows divided into 20 classes, with 10 representing benign applications and 10 representing malware. It reflects real-world traffic, including Skype, Gmail, BitTorrent, and threats such as Zeus and Cridex. With diverse protocols and encrypted content, it serves as a standard benchmark for TC.

PCAP files are processed using Scapy, with flows extracted based on the standard 5-tuple (source/destination IPs and ports, protocol). The Ethernet layer is removed, and IPs are zeroed to prevent overfitting to static identifiers. The resulting byte streams are normalized to [0,1] and stored as byte sequences.

The dataset is processed into two distinct input formats. In the flattened representation, each flow is truncated or padded to 784 bytes, following the convention introduced in~\cite{wang2017end, wang2017malware}. In the time-series format, each flow is segmented into 10 packets, with each packet represented as a 1000-byte vector, producing a $10 \times 1000$ matrix as adopted in~\cite{zhu2023cmtsnn, yao2019identification}.

\subsection{HW-NAS Implementation and Configuration} \label{sec:hwnas_setup}

All experiments were conducted on a workstation equipped with an NVIDIA RTX 3070 Ti GPU, using TensorFlow and Keras for model training. The HW-NAS framework is run separately for the flattened and time-series input formats, using preprocessed flow-level data to search for architectures that maximize validation accuracy under resource constraints.

To define realistic constraints, we analyzed prior malware TC models. When resource metrics were not reported, we estimated them using Keras implementations based on architectural details. This analysis revealed considerable overhead in many designs, motivating stricter limits in our search. Specifically, we constrain architectures to use fewer than \(70\text{k}\) parameters (Flash), under \(3\text{M}\) FLOPs, and a maximum intermediate tensor size below \(6\text{k}\) elements (RAM).

For each input format, a holdout validation set comprising 20\% of the training data is used during search. Architectures are trained for up to 30 epochs using the Adam optimizer with initial learning rate 0.004, batch size 1024 (generator-based), learning rate decay on plateau, and early stopping.

The search runs for 100 generations, each generating 15 candidate architectures via mutation. The search space includes filter counts from 16 to 128, kernel sizes between 2 and 7, strides from 1 to 7, and dropout rates in the range 0.1–0.5. Convolutional padding is selected from \texttt{valid} or \texttt{same}, and pooling layers are either max or average with pool sizes of 2 or 3. The architecture with the highest validation accuracy across all generations is selected as the final model.

\subsection{Adversarial Setup} \label{sec:adv_eval}

Adversarial robustness is evaluated for the final HW-NAS-optimized models on both input formats. Each model, trained on clean data, is tested against FGSM and PGD to assess initial sensitivity. The perturbation magnitude $\epsilon$ is varied in the range $[0.01, 0.2]$ to capture sensitivity under increasing attack strength. FGSM is applied as a single-step $\ell_\infty$ attack. PGD is configured with 10 iterations and a step size of $\alpha = \epsilon / 10$, maintaining the same perturbation budget scaled with $\epsilon$. All attacks are implemented using the Adversarial Robustness Toolbox (ART)~\cite{art_nicolae2018adversarial}, with perturbations applied only to payload bytes. Header fields are masked during gradient computation to preserve structural semantics and prevent unrealistic manipulation, following the constraints used in~\cite{zhang2024adversarial}.

We then apply adversarial fine-tuning to both models. Each is retrained for 100 epochs using FGSM with $\epsilon = 0.1$; this is a common mid-range value that balances robustness and generalization. Training follows a batch-level regime, where every mini-batch consists of 50\% clean and 50\% adversarial samples generated on-the-fly. Afterward, the models are re-evaluated under both FGSM and PGD using the same attack parameters to assess robustness gains.

\section{Results}
\subsection{Resulting Architectures from HW-NAS}

Table~\ref{tab:cnn_architectures_combined} details the final CNN architectures selected by HW-NAS for both input formats. Each row represents a model layer, with columns indicating the layer type, input dimension (Input Dim), number of filters or units (Filt/Units), kernel or pooling size (Ker/Pool), stride (Str), and padding scheme (Pad). The flat-input model stacks four Conv1D layers with progressive compression and early average pooling; the time-series design adopts a shallower structure with two stacked convolutional layers and no pooling.

The flat-input network has 53.02k parameters, 1.99M FLOPs, and a maximum tensor size of 4.88k elements. The time-series model includes 61.45k parameters, 1.18M FLOPs, and a maximum tensor of 1.12k elements. Both models are suitable for edge platforms with at least 256kB of Flash and 20kB of RAM, and scale across typical IoT-class devices.

\begin{table}[htbp]
\centering
\caption{Best CNN architectures produced by HW-NAS}
\label{tab:cnn_architectures_combined}
\renewcommand{\arraystretch}{1.1}
\resizebox{\columnwidth}{!}{%
\begin{tabular}{c|c|c|c|c|c|c}
\hline
\textbf{\#} & \textbf{Layer} & \textbf{Input Dim} & \textbf{Filt/Units} & \textbf{Ker/Pool} & \textbf{Str} & \textbf{Pad} \\
\hline
\multicolumn{7}{c}{\textbf{Flat Input (784)}} \\
\hline
1 & Conv1D + AvgPool & 784$\times$1   & 25  & 7 / 2 & 4 & valid \\
2 & Conv1D           & 98$\times$25   & 90  & 7     & 5 & valid \\
3 & Conv1D           & 19$\times$90   & 70  & 4     & 4 & same \\
4 & Conv1D           & 5$\times$70    & 47  & 3     & 1 & same \\
5 & GAP            & 5$\times$47  & -- & -- & -- & -- \\
6 & Dense + Softmax & 1$\times$47 & 20 & -- & -- & -- \\
\hline
\multicolumn{7}{c}{\textbf{Time-Series Input (10$\times$1000)}} \\
\hline
1 & Conv1D           & 10$\times$1000 & 17  & 3     & 1 & same \\
2 & Conv1D           & 10$\times$1000 & 112 & 4     & 1 & same \\
3 & GAP              & 10$\times$112 & --  & -- & -- & -- \\
4 & Dense + Softmax  & 1$\times$112  & 20  & -- & -- & -- \\
\hline
\end{tabular}
}
\end{table}
\subsection{Clean-Data Performance and Hardware Comparison}

\begin{figure}[htbp]
    \centering
    \includegraphics[width=\columnwidth]{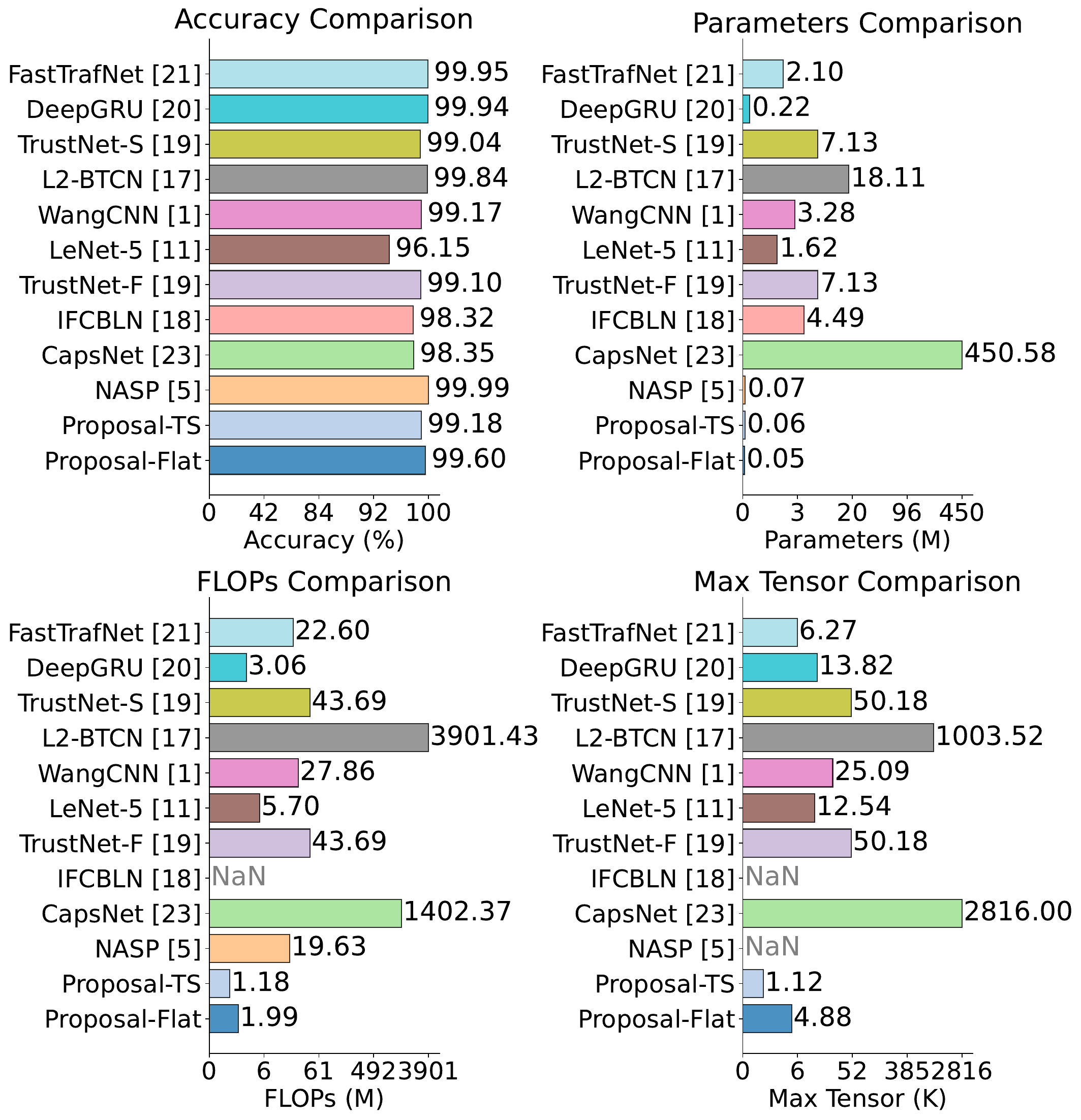}
    \caption{Performance metrics comparison on USTC-TFC2016.}

    \label{fig:clean_results}
\end{figure}

Figure~\ref{fig:clean_results} shows a comparative analysis of clean-data performance across HW-NAS models (denoted as \textit{Proposal-Flat} for flattened flow input and \textit{Proposal-TS} for time-series) and prior state-of-the-art methods. The top-left subplot shows accuracy, while the remaining subplots report model complexity in terms of parameter count, FLOPs, and maximum tensor size.

\vspace{-3mm}

Our HW-NAS models achieve top-tier performance under strict complexity constraints. The flat-input variant achieves 99.60\% accuracy, followed by the time-series version at 99.18\%. In terms of complexity, the time-series model uses 61.45k parameters, slightly more than the flat model’s 53.02k; however, it is more efficient in both FLOPs (1.18M vs. 1.99M) and peak tensor size (1.12k vs. 4.88k). This reflects a trade-off between performance and efficiency: the flat model achieves slightly higher accuracy at the cost of compute and memory; the time-series variant captures temporal structure with lower runtime cost, but requires longer inputs (10$\times$1000 bytes) compared to the flat model’s first 784 bytes of each flow.

Among flow-based baselines, most models process flattened byte streams. NASP~\cite{nas_zhang2023automatic} is a NAS-optimized design that achieves 99.99\% accuracy with 0.07M parameters and 19.63M FLOPs, though intermediate tensor size is not reported. TrustNet-F~\cite{li2025trustworthy} reports 99.10\% using 7.13M parameters and 43.69M FLOPs, while IFCBLN~\cite{xu2024cascaded} yields 98.32\% with 4.49M parameters. In contrast, CapsNet~\cite{yao2019capsule} operates on time-series flows and achieves 98.35\%, but at the cost of 450.58M parameters and over 1400M FLOPs. LeNet-5~\cite{zhang2024adversarial} achieves 96.15\% using 1.62M parameters and 5.7M FLOPs, but falls short in performance compared to other methods.

Session-based methods remain competitive, but are considerably heavier. WangCNN~\cite{wang2017malware} reports 99.17\% with 3.28M parameters and 25.09M FLOPs. L2-BTCN~\cite{li2024l2} reaches 99.84\% but at the cost of 18.11M parameters, 3901.43M FLOPs, and 1003.52k tensors. TrustNet-S~\cite{li2025trustworthy} achieves 99.04\% with 7.13M parameters and 43.69M FLOPs. Packet-level models follow a similar trend: DeepGRU~\cite{ieeeaccess_wang2020deep} and FastTrafNet~\cite{yu2025model} attain 99.94\% and 99.95\%, but require 0.22M and 2.10M parameters, and 3.06M and 22.60M FLOPs, respectively.

Overall, our HW-NAS models provide state-of-the-art accuracy with $<65$k parameters, $<2$M FLOPs, and maximum tensors under 5k. This efficiency is far beyond prior works and makes them well suited for low-power embedded scenarios.

\subsection{Robustness Before Adversarial Training}

Table~\ref{tab:robustness_before_at} reports accuracy (\%) under FGSM and PGD attacks for various perturbation magnitudes $\epsilon$, before adversarial training. Rows group the attack type; columns list increasing $\epsilon$ values from 0.01 to 0.20. Each model’s robustness degrades with stronger perturbations, but at different rates.

The flat-input model shows gradual drops under FGSM, from 99.31\% at $\epsilon=0.01$ to 69.71\% at $\epsilon=0.2$; PGD causes a sharper decline to 48.69\%. The time-series model degrades faster, reaching 32.23\% under FGSM and 23.01\% under PGD at $\epsilon=0.1$, likely due to its larger input (10$\times$1000 bytes) exposing more features to perturbation. 
These results confirm the necessity of targeted defenses, especially when input size increases model vulnerability.

\begin{table}[ht]
\centering
\caption{FGSM/PGD accuracy before adversarial training.}
\huge

\resizebox{\columnwidth}{!}{%
\begin{tabular}{llccccccc}
\toprule
\multirow{2}{*}{Attack} & \multirow{2}{*}{Model} & \multicolumn{7}{c}{$\epsilon$} \\
\cmidrule(lr){3-9}
 &  & 0.01 & 0.03 & 0.05 & 0.07 & 0.10 & 0.15 & 0.20 \\
\midrule
\multirow{3}{*}{FGSM} 
& \textbf{Proposal-Flat} & 99.31 & 97.29 & 93.99 & 91.16 & 86.49 & 76.49 & 69.71 \\
& \textbf{Proposal-TS}   & 89.03 & 50.86 & 40.87 & 36.23 & 32.23 & 27.88 & 25.39 \\
\midrule
\multirow{2}{*}{PGD} 
& \textbf{Proposal-Flat} & 99.27 & 96.22 & 90.03 & 84.39 & 74.78 & 58.47 & 48.69 \\
& \textbf{Proposal-TS}    & 85.82 & 34.60 & 27.26 & 25.01 & 23.01 & 21.45 & 20.73 \\
\bottomrule
\end{tabular}
}
\label{tab:robustness_before_at}
\end{table}

\subsection{Robustness After Adversarial Training}

Table~\ref{tab:robustness_after_at} presents accuracy (\%) under FGSM and PGD attacks after adversarial fine-tuning. The setup mirrors the pre-attack evaluation: rows denote attack type; columns indicate increasing perturbation strength $\epsilon$. All models were trained with FGSM at $\epsilon = 0.1$ and evaluated across the full range.

Adversarial training leads to substantial gains. At $\epsilon = 0.10$, the flat model reaches 96.84\% (FGSM) and 93.20\% (PGD); the time-series model improves more sharply to 88.08\% and 84.37\%, respectively, though it remains more sensitive overall.
Both models also preserve high accuracy on clean data (98.98\% for flat, 98.71\% for time-series). 
These results confirm that adversarial robustness can be boosted in hardware-constrained models, particularly those more sensitive to perturbations, without sacrificing efficiency.


\begin{table}[ht]
\centering
\caption{FGSM/PGD accuracy after adversarial training.}
\huge
\resizebox{\columnwidth}{!}{%
\begin{tabular}{llccccccc}
\toprule
\multirow{2}{*}{Attack} & \multirow{2}{*}{Model} & \multicolumn{7}{c}{$\epsilon$} \\
\cmidrule(lr){3-9}
 &  & 0.01 & 0.03 & 0.05 & 0.07 & 0.10 & 0.15 & 0.20 \\
\midrule
\multirow{3}{*}{FGSM} 
& \textbf{Proposal-Flat} & 98.65 & 97.95 & 97.45 & 97.42 & 96.84 & 95.50 & 89.86 \\
& \textbf{Proposal-TS}    & 98.32 & 97.00 & 94.95 & 92.19 & 88.08 & 83.63 & 76.80 \\
\midrule
\multirow{2}{*}{PGD} 
& \textbf{Proposal-Flat} & 98.62 & 97.54 & 96.55 & 95.19 & 93.20 & 87.94 & 78.27 \\
& \textbf{Proposal-TS}    & 98.31 & 96.65 & 94.07 & 90.34 & 84.37 & 73.29 & 59.72 \\
\bottomrule
\end{tabular}
}
\label{tab:robustness_after_at}
\end{table}


\section{Conclusion}

This work explored the interplay between hardware efficiency, input representation, and adversarial robustness in TC. Using HW-NAS, we derived compact DNNs that deliver high accuracy under strict resource constraints, across the two dominant input formats in TC: flattened byte sequences and time-series packet matrices. Post-hoc adversarial analysis revealed distinct vulnerability profiles, with the flat-input model exhibiting stronger resilience. Batch-level adversarial fine-tuning further boosted robustness across perturbation levels. The resulting architectures are deployable across a wide range of IoT platforms. Future work will explore adaptive and black-box defenses to improve resilience under broader threat models, and extend evaluation to diverse tasks and datasets.


\begin{thebibliography}{10}
\providecommand{\url}[1]{#1}
\csname url@samestyle\endcsname
\providecommand{\newblock}{\relax}
\providecommand{\bibinfo}[2]{#2}
\providecommand{\BIBentrySTDinterwordspacing}{\spaceskip=0pt\relax}
\providecommand{\BIBentryALTinterwordstretchfactor}{4}
\providecommand{\BIBentryALTinterwordspacing}{\spaceskip=\fontdimen2\font plus
\BIBentryALTinterwordstretchfactor\fontdimen3\font minus \fontdimen4\font\relax}
\providecommand{\BIBforeignlanguage}[2]{{%
\expandafter\ifx\csname l@#1\endcsname\relax
\typeout{** WARNING: IEEEtran.bst: No hyphenation pattern has been}%
\typeout{** loaded for the language `#1'. Using the pattern for}%
\typeout{** the default language instead.}%
\else
\language=\csname l@#1\endcsname
\fi
#2}}
\providecommand{\BIBdecl}{\relax}
\BIBdecl

\bibitem{wang2017malware}
W.~Wang, M.~Zhu, X.~Zeng, X.~Ye, and Y.~Sheng, ``Malware traffic classification using convolutional neural network for representation learning,'' in \emph{2017 International conference on information networking (ICOIN)}.\hskip 1em plus 0.5em minus 0.4em\relax IEEE, 2017, pp. 712--717.

\bibitem{zhu2023cmtsnn}
S.~Zhu, X.~Xu, H.~Gao, and F.~Xiao, ``Cmtsnn: A deep learning model for multiclassification of abnormal and encrypted traffic of internet of things,'' \emph{IEEE Internet of Things Journal}, vol.~10, no.~13, pp. 11\,773--11\,791, 2023.

\bibitem{survey_DONG2024128444}
W.~Dong, J.~Yu, X.~Lin, G.~Gou, and G.~Xiong, ``Deep learning and pre-training technology for encrypted traffic classification: A comprehensive review,'' \emph{Neurocomputing}, p. 128444, 2024.

\bibitem{wang2017end}
W.~Wang, M.~Zhu, J.~Wang, X.~Zeng, and Z.~Yang, ``End-to-end encrypted traffic classification with one-dimensional convolution neural networks,'' in \emph{2017 IEEE international conference on intelligence and security informatics (ISI)}.\hskip 1em plus 0.5em minus 0.4em\relax IEEE, 2017, pp. 43--48.

\bibitem{nas_zhang2023automatic}
X.~Zhang, L.~Hao, G.~Gui, Y.~Wang, B.~Adebisi, and H.~Sari, ``An automatic and efficient malware traffic classification method for secure internet of things,'' \emph{IEEE Internet of Things Journal}, vol.~11, no.~5, pp. 8448--8458, 2023.

\bibitem{our_chehade2025iscc}
A.~Chehade, E.~Ragusa, P.~Gastaldo, and R.~Zunino, ``Energy-efficient deep learning for traffic classification on microcontrollers,'' \emph{arXiv preprint arXiv:2506.10851}, 2025.

\bibitem{ragusa2024combining}
E.~Ragusa, F.~Zonzini, P.~Gastaldo, and L.~De~Marchi, ``Combining compressed sensing and neural architecture search for sensor-near vibration diagnostics,'' \emph{IEEE Transactions on Industrial Informatics}, 2024.

\bibitem{our_chehade2025tnsm}
A.~Chehade, E.~Ragusa, P.~Gastaldo, and R.~Zunino, ``Efficient traffic classification using hw-nas: Advanced analysis and optimization for cybersecurity on resource-constrained devices,'' \emph{arXiv preprint arXiv:2506.11319}, 2025.

\bibitem{roshan2024untargeted}
K.~Roshan, A.~Zafar, and S.~B.~U. Haque, ``Untargeted white-box adversarial attack with heuristic defence methods in real-time deep learning based network intrusion detection system,'' \emph{Computer Communications}, vol. 218, pp. 97--113, 2024.

\bibitem{alhussien2024constraining}
N.~Alhussien, A.~Aleroud, A.~Melhem, and S.~Y. Khamaiseh, ``Constraining adversarial attacks on network intrusion detection systems: transferability and defense analysis,'' \emph{IEEE Transactions on Network and Service Management}, vol.~21, no.~3, pp. 2751--2772, 2024.

\bibitem{zhang2024adversarial}
C.~Zhang and P.~Wang, ``Adversarial attack defense algorithm based on convolutional neural network,'' \emph{Neural Computing and Applications}, vol.~36, no.~17, pp. 9723--9735, 2024.

\bibitem{fgsm_goodfellow2014explaining}
I.~J. Goodfellow, J.~Shlens, and C.~Szegedy, ``Explaining and harnessing adversarial examples,'' \emph{arXiv preprint arXiv:1412.6572}, 2014.

\bibitem{pgd_madry2017towards}
A.~Madry, A.~Makelov, L.~Schmidt, D.~Tsipras, and A.~Vladu, ``Towards deep learning models resistant to adversarial attacks,'' \emph{arXiv preprint arXiv:1706.06083}, 2017.

\bibitem{ports_niu2019heuristic}
W.~Niu, Z.~Zhuo, X.~Zhang, X.~Du, G.~Yang, and M.~Guizani, ``A heuristic statistical testing based approach for encrypted network traffic identification,'' \emph{IEEE Transactions on Vehicular Technology}, vol.~68, no.~4, pp. 3843--3853, 2019.

\bibitem{signature_dpi1_wang2020automatic}
X.~Wang, S.~Chen, and J.~Su, ``Automatic mobile app identification from encrypted traffic with hybrid neural networks,'' \emph{Ieee Access}, vol.~8, pp. 182\,065--182\,077, 2020.

\bibitem{dpi_hongke2022dpi}
\BIBentryALTinterwordspacing
{HongKe}. (2022) Hongke sharing | what is deep packet inspection (dpi)? (chinese). (accessed 09 October 2024). [Online]. Available: \url{https://zhuanlan.zhihu.com/p/572823255}
\BIBentrySTDinterwordspacing

\bibitem{li2024l2}
Z.~Li and X.~Xu, ``L2-bitcn-cnn: Spatio-temporal features fusion-based multi-classification model for various internet applications identification,'' \emph{Computer Networks}, vol. 243, p. 110298, 2024.

\bibitem{xu2024cascaded}
J.~Xu, Y.~Zhang, K.~Zhou, Q.~Wang, M.~Hua, L.~Shan, Y.~Lin, and G.~Gui, ``A cascaded broad learning network embedded image features for malware traffic classification,'' \emph{IEEE Transactions on Cognitive Communications and Networking}, 2024.

\bibitem{li2025trustworthy}
Z.~Li, Y.~Liu, C.~Zhang, W.~Shan, H.~Zhang, and X.~Zhu, ``Trustworthy deep learning for encrypted traffic classification,'' \emph{Soft Computing}, pp. 1--18, 2025.

\bibitem{ieeeaccess_wang2020deep}
B.~Wang, Y.~Su, M.~Zhang, and J.~Nie, ``A deep hierarchical network for packet-level malicious traffic detection,'' \emph{IEEE Access}, vol.~8, pp. 201\,728--201\,740, 2020.

\bibitem{yu2025model}
L.~Yu, J.~Yuan, J.~Zheng, and N.~Yang, ``A model of encrypted network traffic classification that trades off accuracy and efficiency,'' \emph{Journal of Network and Systems Management}, vol.~33, no.~1, pp. 1--32, 2025.

\bibitem{yao2019identification}
H.~Yao, C.~Liu, P.~Zhang, S.~Wu, C.~Jiang, and S.~Yu, ``Identification of encrypted traffic through attention mechanism based long short term memory,'' \emph{IEEE trans. big data}, vol.~8, no.~1, pp. 241--252, 2019.

\bibitem{yao2019capsule}
H.~Yao, P.~Gao, J.~Wang, P.~Zhang, C.~Jiang, and Z.~Han, ``Capsule network assisted iot traffic classification mechanism for smart cities,'' \emph{IEEE Internet of Things Journal}, vol.~6, no.~5, pp. 7515--7525, 2019.

\bibitem{1000papersNAS}
C.~White, M.~Safari, R.~Sukthanker, B.~Ru, T.~Elsken, A.~Zela, D.~Dey, and F.~Hutter, ``Neural architecture search: Insights from 1000 papers,'' \emph{arXiv preprint arXiv:2301.08727}, 2023.

\bibitem{chitty2023neural_ieee-access}
K.~T. Chitty-Venkata, M.~Emani, V.~Vishwanath, and A.~K. Somani, ``Neural architecture search benchmarks: Insights and survey,'' \emph{IEEE Access}, vol.~11, pp. 25\,217--25\,236, 2023.

\bibitem{ragusa2024compression}
E.~Ragusa, F.~Zonzini, L.~De~Marchi, and R.~Zunino, ``Compression-accuracy co-optimization through hardware-aware neural architecture search for vibration damage detection,'' \emph{IEEE Internet of Things J.}, 2024.

\bibitem{our_chehade2024tiny}
A.~Chehade, E.~Ragusa, P.~Gastaldo, and R.~Zunino, ``Tiny neural networks for session-level traffic classification,'' in \emph{International Conference on Applications in Electronics Pervading Industry, Environment and Society}.\hskip 1em plus 0.5em minus 0.4em\relax Springer, 2024, pp. 347--354.

\bibitem{maarouf2021evaluating}
R.~Maarouf, D.~Sattar, and A.~Matrawy, ``Evaluating resilience of encrypted traffic classification against adversarial evasion attacks,'' in \emph{2021 IEEE Symposium on Computers and Communications (ISCC)}.\hskip 1em plus 0.5em minus 0.4em\relax IEEE, 2021, pp. 1--6.

\bibitem{sadeghzadeh2021adversarial}
A.~M. Sadeghzadeh, S.~Shiravi, and R.~Jalili, ``Adversarial network traffic: Towards evaluating the robustness of deep-learning-based network traffic classification,'' \emph{IEEE Transactions on Network and Service Management}, vol.~18, no.~2, pp. 1962--1976, 2021.

\bibitem{art_nicolae2018adversarial}
M.-I. Nicolae, M.~Sinn, M.~N. Tran, B.~Buesser, A.~Rawat, M.~Wistuba, V.~Zantedeschi, N.~Baracaldo, B.~Chen, H.~Ludwig \emph{et~al.}, ``Adversarial robustness toolbox v1. 0.0,'' \emph{arXiv preprint arXiv:1807.01069}, 2018.

\end{thebibliography}

\end{document}